\documentclass[twocolumn,prc,showpacs,amsmath,amssymb]{revtex4-1}

\usepackage{graphicx}

\usepackage{soul}
\usepackage{calrsfs}
\usepackage{pgfplotstable}
\usepackage{pgfplots}
\usepackage{multirow}

\usetikzlibrary{patterns}
\usepgfplotslibrary{groupplots}

\newcommand{\CM}{\mathcal{M}}
\newcommand{\CA}{\mathcal{A}}
\newcommand{\CF}{\mathcal{F}}

\usepackage{color}

\usetikzlibrary{external}

\begin{document}

\long\def\/*#1*/{}

\title{Magnetic moments in odd-A Cd isotopes and
  coupling of particles with the zero-point vibrations}

\author{S. Mishev}
\email{mishev@theor.jinr.ru}
\affiliation{Joint Institute for Nuclear Research, 6 Joliot-Curie str., Dubna 141980, Russia}
\affiliation{Institute for Advanced Physical Studies, New Bulgarian University, 21 Montevideo str., Sofia 1618, Bulgaria}
\author{V.V. Voronov}
\affiliation{Joint Institute for Nuclear Research, 6 Joliot-Curie str., Dubna 141980, Russia}
\date{\today}

\begin{abstract}
\begin{description}
\item[Background] The coupling of the last nucleon with configurations
  in the ground state of the even-even core is known to augment the
  single quasiparticle fragmentation pattern. In a recent experimental
  study by Yordanov \emph{et al.} the values of the magnetic dipole
  and electric quadrupole moments of the $11/2^-$ state in a long
  chain of Cd isotopes were found to follow a simple trend which we
  try to explain by means of incorporating long-range correlations
  in the ground state.
\item[Purpose] Our purpose is to study the influence of the
  ground-state correlations (GSC) on the magnetic moments and compare our
  results with the data for the odd-A Cd isotopes.
\item[Method] In order to evaluate if the additional correlations have
  bearing on the magnetic moments we employ an extension to the
  quasiparticle-phonon model (QPM) which takes into account
  quasiparticle$\otimes$phonon configurations in the ground state of
  the even-even core to the structure of the odd-A nucleus wave
  function.
\item[Results] It is shown that the values for the magnetic moments
  which the applied QPM extension yields deviate further from the Schmidt
  values. The latter is in agreement with the measured values for the Cd
  isotopes.
\item[Conclusions] The GSC exert significant influence on the magnetic
  dipole moments and reveal a potential for reproducing the
  experimental values for the studied cadmium isotopes.
\end{description}
\end{abstract}



\maketitle

\section{Introduction}

The important role that the particle-vibration interaction plays in 
explaining the deviations of nuclear magnetic moments from the Schmidt
values was first acknowledged in Refs. \cite{1954_blin-stoyle,
  1954_arima}. The strength of this interaction depends mainly on the
state which the participating nucleon occupies as well as on the
distribution of multiparticle--multihole configurations constituting
the vibrating core. Intuitively clear, still supported by the
experimental data, is the finding that this interaction is weak near
the magic nuclei and becomes substantial in the open-shell
regions. For example, performing calculations relying on the first and
second order perturbation theory, as shown in Ref.\cite{1986_arima},
this interaction manifests itself as capable of explaining the
differences between the magnetic moments in the near magic odd-A Tl
isotopes. In the open-shell regions, however, there is a large number
of nuclei, referred to as transitional, in which the diversity of
configurations contributing to the vibration grows rapidly and the
terms which could be neglected near the magic configurations are no
longer small. In this respect the data and analysis by Yordanov
et.al. \cite{2013_yordanov}, which we interpret in this work, are
important for at least two reasons. In the first place, the long chain
of cadmium isotopes, which is explored in Ref.\cite{2013_yordanov},
enters the transitional region with respect to the neutron
subsystem. Secondly, the measured quantities, namely, the magnetic
dipole and electric quadrupole moments, not only triggered the
invention of the nuclear shell model and the collective
model\cite{1950_mayer, 1976_bohr} but still are a major tool to test
the validity of modern nuclear theories.

The theoretical interpretations of the data concerning the quadrupole
moments, reported in Ref.\cite{2013_yordanov}, were previously carried
out by using two kinds of pairing models - the seniority scheme
\cite{1963_shalit} and the BCS approximation\cite{1957_bcs}. Decent
agreement with the experimental data is reached by the authors of
Ref. \cite{2013_yordanov} using the seniority model at the cost of
introducing an effective neutron charge and by neglecting
configurations with seniorities greater than 1. A more robust study,
which does not compromise the use of effective charges, is performed in
Ref.\cite{2014_Zhao} by employing the BCS approximation to account for
the pairing of nucleons. The important result of this study is the
estimation that the contribution from the polarized core with 40
protons to the quadrupole moment is as large as the contribution from
the valence protons in the g9/2 orbital. The authors of
Ref. \cite{2011_tolokonnikov} elaborated on the influence of the
density dependence of the effective pairing interaction on the
quadrupole moments in odd-N Sn and Pb nuclei, which they found to be
notable and produce deviations between 10$\%$ and 50$\%$ in different
isotopes. The electric quadrupole moments in odd-A Sn nuclei have been
successfully reproduced in Ref. \cite{2014_Jiang} using the
nucleon-pair approximation of the shell model, and also the magnetic
moments were found to be very close to their single-particle
estimates.

If the pairing correlations seem sufficient to describe the trends in
the behavior of the quadrupole moments, the magnetic moments of the
low-lying states require further efforts mainly due to the role of the
$1^{+}$ magnetic excitations and the nucleon correlations in the
ground state induced by the long-range part of the residual
nucleon-nucleon force. Although the contribution to the wave function
coming from configurations owing to the coupling with the magnetic
giant dipole resonance are small their influence on the magnetic
moments is significant because of the strong M1 transition to the
ground state. In Ref. \cite{2008_borzov} a systematic theoretical
analysis of experimental data on magnetic moments in different nuclei
is performed utilizing the theory of finite Fermi systems. The effects
of the M1 giant resonance on the magnetic moments (the Arima-Horie
effect) were the main focus of many research studies while the
contribution from other modes seem to be a less explored
territory. The configuration mixing generated by such modes can be
taken into account in the calculations of the magnetic moments by
either introducing effective two-body operators \cite{1971_Zamick,
  1972_Nomura} or evaluating the average values of the single-particle
operator in multiparticle configurations.

The influence on the magnetic moments coming from the coupling of the
last nucleon with the low-lying collective quadrupole and octupole
core excitations is studied in Ref. \cite{1986_levon}. The generated
admixtures from these interactions were found to give important
contributions in most of the studied isotopes but are most considerable in
odd-Z nuclei.

In the present development we account for the above effects relying on
the concepts and instruments of the quasiparticle phonon model (QPM)
\cite{1992_soloviev} without making use of two-body operators for the
magnetic moment. Of special interest for our research is the evaluation
of the role of the quasiparticle and the quasiparticle$\otimes$phonon
configurations residing in the ground state of the even-even core on
the magnetic moments. This problem, which has not been explored so
far, is approached by allowing non zero values for the amplitudes in
the wave function of the odd-A nucleus corresponding to the above
configurations \cite{1965_Kuo, 1993_Sluys, 2008_mishev,
  2010_mishev}. Our previous investigation on this topic showed
considerable fragmentation of the low-energy single-quasiparticle
states due to such long-range correlations. The latter is a crux in
reliably applying the perturbation theory, which otherwise yields
rather tractable results \cite{1954_blin-stoyle, 1954_arima,
  1986_arima, 2014_Tekacsy}. In applying this idea using the so-called
backward-going amplitudes in the odd-A nucleus wave function to
calculate the expectation value of the magnetic operator we registered
considerable shifts in the magnetic moments. The subtle effects owing
to the account of the ground state correlations is what distinguishes
our work from the studies performed in Ref. \cite{1986_levon}.

The meson exchange currents between the nucleons inside a nucleus,
which modify the single-particle nature of the magnetic moment
operator \cite{1951_miyazawa}, are taken into account by
introducing effective $g_s$-factors values.

This paper proceeds as follows. In Sec. \ref{sec:th_framework}, we
outline the basics of the approach that we utilize to estimate the
magnetic moments. Numerics and physical interpretations are the
subject of Sec. \ref{sec:results}. The results of this work are
summarized in Sec. \ref{sec:outlook}.

\section{Theoretical framework}
\label{sec:th_framework}

In the approach that we follow, the properties of the nuclear states
are interpreted as a result of the interaction between two types of
fictitious particles - quasiparticles and phonons represented in
spherical basis by operators denoted by $\alpha_{jm}$ and
$Q_{\lambda\mu i}$, respectively \cite{1992_soloviev}. In this
framework the odd-even nuclei are formed by the interaction of the
last quasiparticle with the ground and excited states of the even-even
core. The possibilities for the last particle to couple with different
states of the even-even core are accounted for by constructing a wave
function as a mixture of one quasiparticle and quasiparticle-phonon
pure states \cite{2008_mishev,2010_mishev}:

\begin{align}
  \label{eq:wv_odd_even_bcw}
  \Psi_\nu (JM) = O^{\dagger}_{JM\nu} |\rangle
\end{align}

with

\begin{align}
  \label{eq:oper_odd_even_bcw}
  O^{\dagger}_{JM\nu} = & C_{J\nu}\alpha _{JM}^\dagger +  \sum_{j\lambda i}  D_{j\lambda i}(J\nu )P_{j\lambda i}^{\dagger}(JM) - \nonumber \\
  &E_{J\nu}\tilde{{\alpha}}_{JM}-\sum_{j\lambda i} F_{j\lambda i}(J\nu)\tilde{P}_{j\lambda i}(JM),
\end{align}

where $|\rangle$ denotes the ground state of the even-even core,
$P^{\dagger}(JM) = [\alpha^{\dagger}_{j}Q^{\dagger}_{\lambda i}]_{JM}$
is the quasiparticle$\otimes$phonon creation operator and $\tilde{}$
stands for time conjugation according to the convention
$\tilde{a}_{jm} = (-1)^{j-m}a_{j-m}$. The last terms of this equation
address the non-zero probabilities for the last quasiparticle to
interact with quasiparticle and quasiparticle$\otimes$phonon
configurations residing in the ground state of the even-even core. The
importance of these terms to the magnetic and electric moments is
discussed in detail in Sec. \ref{sec:results}.

The dynamics of the physical setting described in this way is governed
by the following Hamiltonian :

\begin{equation}
  H = H_{MF} + H_{PAIR} + H_{RES},
\end{equation}

which includes a part representing the integral effect from the
mean-field generating forces of the nucleon-nucleon interaction,
monopole pairing field and the residual central long-range interaction
between the spatial and spin degrees of freedom:

\begin{equation}
  \label{eq:interaction}
  H_{RES} = H_{M} + H_{SM}.
\end{equation}

Assuming this part of the interaction in a separable form we expand it
by multipoles and spin-multipoles:

\begin{equation}
  \label{eq:interaction_multipole-multipole}
  H_{M} = -\frac 12 \sum_{\substack{\lambda \\ \rho=\pm 1 }}(\kappa_{0}^{(\lambda)}+\rho\kappa_{1}^{(\lambda)})\sum_{\substack{\mu \\ \tau=n,p }}M^{\dagger}_{\lambda\mu}(\tau)M_{\lambda\mu}(\rho\tau),
\end{equation}

\begin{equation}
  \label{eq:interaction_spin-multipole}
  H_{SM} = -\frac 12 \sum_{\substack{\lambda \\ L=\lambda, \lambda \pm 1 \\ \rho=\pm 1 }}(\kappa_{0}^{(\lambda L)} + \rho \kappa_{1}^{(\lambda L)})\sum_{\substack{M \\ \tau=n,p }}(S^{\lambda}_{LM})^{\dagger}(\tau)S^{\lambda}_{LM}(\rho\tau),
\end{equation}

where $\tau$ enumerates the neutron ($n$) and proton ($p$) subsystems.

\begin{equation}
  \label{def:M}
  M^{\dagger}_{\lambda\mu} = \sum_{jj^{\prime}mm^{\prime}} \langle jm | i^{\lambda} R_{\lambda}(r)Y_{\lambda\mu}  | j^{\prime}m^{\prime} \rangle a^{\dagger}_{j^{\prime}m^{\prime}} a_{jm} 
\end{equation}

and

\begin{equation}
  \label{def:S}
  (S^{\lambda}_{LM})^{\dagger} = \sum_{jj^{\prime}mm^{\prime}} \langle jm | i^{\lambda}R_{\lambda}(r)[\sigma Y_{\lambda}]_{LM} | j^{\prime}m^{\prime} \rangle a^{\dagger}_{j^{\prime}m^{\prime}} a_{jm}
\end{equation}

are the single-particle multipole and spin-multipole operators
\cite{1992_soloviev}. From the sum over $L$ in
Eq. \eqref{eq:interaction_spin-multipole} we include only the terms
with $L=\lambda - 1$. In the phonon space, the eigenstates
of this part of the interaction are of unnatural parity
$(-1)^{L-1}$. Of particular interest to our present research are the
$1^{+}$ states which in QPM account for the dipole core spin
polarization \cite{1986_levon}, induced by the $\sigma\sigma$ forces. The
reduced matrix elements related to equations \eqref{def:M} and
\eqref{def:S} are denoted by $f^{(\lambda)}_{jj'}$ and $f^{(\lambda L)}_{jj'}$ respectively.

We obtain the eigenstates of the system, defined by the Hamiltonian
\eqref{eq:interaction}, using an approximate step-by-step
diagonalization procedure in which initially the first two terms are
diagonalized using the canonical Bogoliubov transformation

\begin{equation}
  \label{eq:def_qp}
  a _{jm} = u_j \alpha_{jm} + (-)^{j-m}v_j \alpha_{j-m}^{\dagger}.
\end{equation}

The term of the
Hamiltonian ($H_{RES}$), which contains non-diagonal elements in the
quasiparticle basis after the first step of this procedure, couples
different quasiparticles to form mixed states which in the QPM are
understood using the concept of phonons

\begin{equation}
  \label{eq:def_phonon}
  Q_{\lambda \mu i}^{\dagger}=\frac 12\sum_{jj^\prime
  }\left[\psi_{jj^\prime}^{\lambda i}\,A^{\dagger}(jj^\prime;\lambda
    \mu)-(-1)^{\lambda -\mu }\varphi _{jj^\prime}^{\lambda
      i}\,A(jj^\prime ;\lambda -\mu )\right],
\end{equation}

, where $A^{\dagger}$ (and its inverse) stand for the bifermion quasiparticle
operator:

\begin{equation}
  \label{eq:oper_A}
  A^{\dagger}(jj^\prime;\lambda \mu )=\sum_{mm^\prime }\langle
  jmj^\prime m^\prime \mid \lambda \mu \rangle \alpha
  _{jm}^{\dagger}\alpha _{j^\prime m^\prime }^{\dagger}.
\end{equation}

In order to determine the structure of even-even nuclei, this part of
the interaction is diagonalized in a space spanned by one-phonon wave
functions where it is assumed that the ground state of the even-even
core is a vacuum for the phonon operators, i.e. $Q_{\lambda\mu i}|
\rangle = 0$. Of special attention is the fact that the core's
ground state from Eq. \eqref{eq:wv_odd_even_bcw} contains additional
correlations that are not included in the phonon vacuum state. The
latter are incorporated by equating numbers rather than wave functions
as theorized in the equation-of-motion method \cite{1970_rowe} which
we apply in the following form:

\begin{equation}
  \label{eq:eom}
  \langle|\{\delta O_{JM\nu }, H, O_{JM\nu }^{\dagger} \}|\rangle = \eta_{J\nu}\langle|\{\delta O_{JM}, O_{JM}^{\dagger} \}|\rangle.
\end{equation}

This method allows to harness the already obtained phonon vacuum state
for calculating the average values of the Hamiltonian. In
Eq.\eqref{eq:eom} $\left\lbrace \cdot , \cdot , \cdot \right\rbrace$
stands for the double commutator and $\eta_{J\nu}$ is the energy of
the $\nu$-th eigenstate with angular momentum $J$. Despite lowering
of the particle rank, the double commutator yields two-body operators
whose average values still depend on the ground-state correlations. In
this work, we evaluate the operators' average values using the random
phase approximation (RPA) with corrections for certain
three-quasiparticle configurations affected by the Pauli exclusion
principle \cite{1981_khuong}.

Having determined the structural composition of the odd-even nucleus,
estimates for the observable quantities of interest are obtained by
evaluating the average values of the corresponding operators. For the
magnetic dipole and electric quadrupole moments they are defined as

\begin{equation}
  \label{eq:magnetic_dipole_moment}
  \mu_1(J\nu) = \sqrt{\frac{4\pi}{3}}  \langle J J \nu| \CM(M;10) | J J \nu\rangle ,
\end{equation}

\begin{equation}
  \label{eq:electric_quadrupole_moment}
  Q_2(J\nu) = \sqrt{\frac{16\pi}{5}} \langle JJ\nu| \CM(E;20) |JJ\nu\rangle ,
\end{equation}

where the electric and magnetic multipole operators are expressed as

\begin{align}
  \label{eq:em_operator}
  & \CM(X;\lambda \mu) = \\
  &\frac{1}{\pi_\lambda }\sum_{\substack{j_1 m_1 \\j_2 m_2 }} {\left( -1 \right)^{j_2-m_2 } \CF_{j_1 j_2 }^{(\lambda)}  \left\langle j_1 m_1 ,j_2  - m_2 | \lambda \mu  \right\rangle a_{j_1 m_1 }^ +  a_{j_2 m_2 } } \nonumber.
\end{align}

Hereafter $\pi_\lambda = \sqrt{2\lambda + 1}$.
$\CF_{j_{1}j_{2}}^{(\lambda)}$ are the reduced single particle matrix
elements:

\begin{widetext}

\begin{equation}
 \CF_{j_{1}j_{2}}^{(\lambda)} = \begin{cases}
   e\left\langle j_{2}|| r^{\lambda}i^{\lambda}Y_{\lambda\mu}  || j_{1} \right\rangle, \mbox{for electric transitions,}\\
   \mu_{0} \left( g_{s} \left\langle j_{2}|| s. \nabla (r^{\lambda} Y_{\lambda\mu}) || j_{1} \right\rangle + g_{l} \frac{2}{\lambda+1}\left\langle j_{2}|| l. \nabla (r^{\lambda} Y_{\lambda\mu}) || j_{1} \right\rangle \right), \mbox{for magnetic triansitions}.
 \end{cases}
\end{equation}

\end{widetext}

Here $e$ and $\mu_0$ are the electron charge and nuclear magneton,
respectively.

The matrix elements of the electromagnetic operator
\eqref{eq:em_operator} in nuclear wave functions derived by using the
independent-particle approximation can be decomposed into two parts
\cite{1981_ring_schuck} - one evaluating its expectation value in the
even-even core and the other representing the matrix element of this
operator between the corresponding single-particle states. For the magnetic moments,
only the second term gives a contribution. Depending on the
degree of correlation of the core in its ground state this simple
picture gets modified by correcting these terms and also by considering
additional terms which vanish in the single-particle model. To take
such corrections into account, we represent all respective quantities
in terms of quasiparticles \eqref{eq:def_qp} and phonons
\eqref{eq:def_phonon}. In that way the matrix elements in Eqs.
\eqref{eq:magnetic_dipole_moment} and
\eqref{eq:electric_quadrupole_moment} are obtained in a form which can
be derived from the following formulae:

\begin{widetext}
  \begin{equation}
    \label{eq:x-qp-qp}
    \langle J_2 M_2 \nu_2| \CM(X;\lambda\mu) | J_1 M_1 \nu_1\rangle = \frac{\langle J_{1}-M_{1} \lambda -\mu|J_{2}-M_{2} \rangle}{\pi_{J_{2}}} (x_{qp-qp} + x_{qp-ph}+ x_{ph-ph} )
\end{equation}

, where $x_{qp-qp}$ gives the transition amplitude between two quasiparticle states

\begin{align}
\label{eq:a-qp-qp}
   x_{qp-qp} = & \left\langle \left| \left\lbrace \left( C_{J_2\nu_{2}}\alpha_{J_2M_2} - E_{J_{2}\nu_{2}}\tilde{\alpha}^{\dagger}_{J_2M_2}  \right), \left[\CM (X;\lambda\mu), C_{J_{1}\nu_{1}}\alpha^{\dagger}_{J_1M_1} - E_{J_{1}\nu_{1}}\tilde{\alpha}_{J_1M_1} \right] \right\rbrace \right|\right\rangle = \nonumber \\
  & = \left( - C_{J_1\nu_{1}} C_{J_2\nu_{2}}   +   E_{J_{1}\nu_{1}} E_{J_{2}\nu_{2}} \right)  \CF_{J_{1}J_{2}}^{\lambda}v^{\pm}_{J_{1}J_{2}}
\end{align}

, $x_{qp-ph}$ evaluates the transition amplitudes between quasiparticle and quasiparticle$\otimes$phonon states :

\begin{align}
  \label{eq:a-qp-ph}
  & x_{qp-ph} = \frac 12 \frac{1}{\pi_{\lambda}} \sum_i \left[ \pi_{J_1} \left( -C_{J_2\nu_2}D_{J_2}^{\lambda i}(J_1\nu_1) + E_{J_2\nu_2}F_{J_2}^{\lambda i}(J_1\nu_1) \right) - (-1)^{J_1+J_2+\lambda} \pi_{J_2} \left( C_{J_1\nu_1}D_{J_1}^{\lambda i}(J_2\nu_2)  -  E_{J_1\nu_1}F_{J_1}^{\lambda i}(J_2\nu_2) \right) \right]\nonumber \\ 
  &\left[\sum_{12}^n \CF_{j_1j_2}^{(\lambda)} u_{j_1j_2}^\mp(\psi_{j_1j_2}^{\lambda i} \mp \varphi_{j_1j_2}^{\lambda i}) + \sum_{12}^p \CF_{j_1j_2}^{(\lambda)} u_{j_1j_2}^\mp(\psi_{j_1j_2}^{\lambda i} \mp \varphi_{j_1j_{2}}^{\lambda i})\right]
\end{align}

and $x_{ph-ph}$ corresponds to the transition amplitudes between two quasiparticle$\otimes$phonon states

\begin{align}
\label{eq:a-ph-ph}
  x_{ph-ph} = &  \pi_{J_1}\pi_{J_2}\sum_{j_1 j_2\lambda' i'} (-)^{j_1+J_2+\lambda'} \left( D_{j_1}^{\lambda' i'}(J_2\nu_2) D_{j_2}^{\lambda' i'}(J_1\nu_1) - F_{j_1}^{\lambda' i'}(J_2\nu_2) F_{j_2}^{\lambda' i'}(J_1\nu_1) \right) \left\{ {\begin{array}{*{20}c}
        {J_1 } & {\lambda } & {J_2 }  \\
        {j_1 } & {\lambda' } & {j_2 }  \\
      \end{array}} \right\} \CF_{j_1j_2}^{(\lambda)} v_{j_1j_2}^\pm.
\end{align}

\end{widetext}

In formulae \eqref{eq:a-qp-qp} - \eqref{eq:a-ph-ph}, the plus and
minus signs in $\pm$ and $\mp$ apply to the magnetic and the electric
moment, respectively. If the residual interaction is switched off,
then the terms \eqref{eq:a-qp-ph} and \eqref{eq:a-ph-ph} disappear and
the term \eqref{eq:a-qp-qp} gives the well-known Schmidt values. The
terms \eqref{eq:a-qp-ph} and \eqref{eq:a-ph-ph} have the following
important difference: while the term \eqref{eq:a-qp-ph} involves
properties of the phonons whose angular momenta coincide with the
multipolarity of the transition operator, the summation in
Eq.\eqref{eq:a-ph-ph}, in contrast, runs over all angular momenta
$\lambda=1,2,3\ldots$.

\section{Results}
\label{sec:results}

\begin{figure}[t]
  \includegraphics{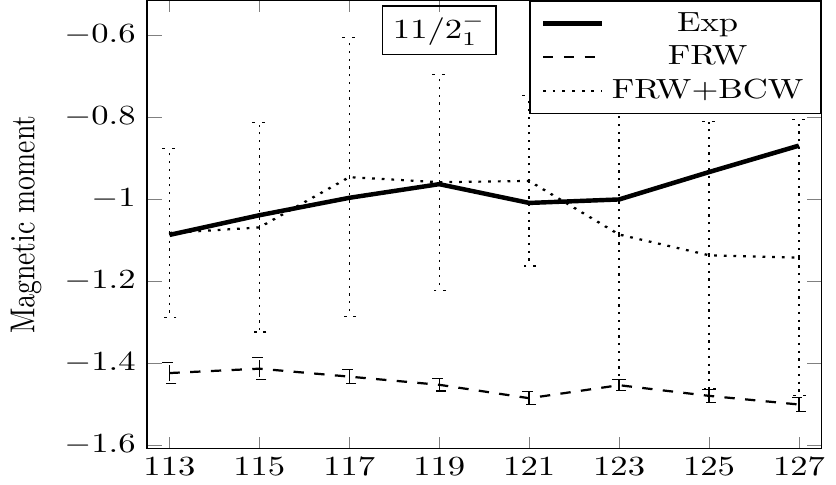}
  \includegraphics{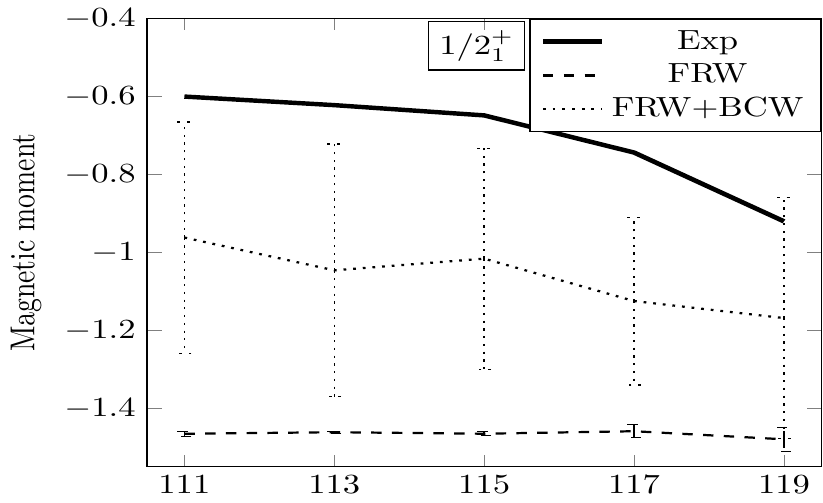}
  \includegraphics{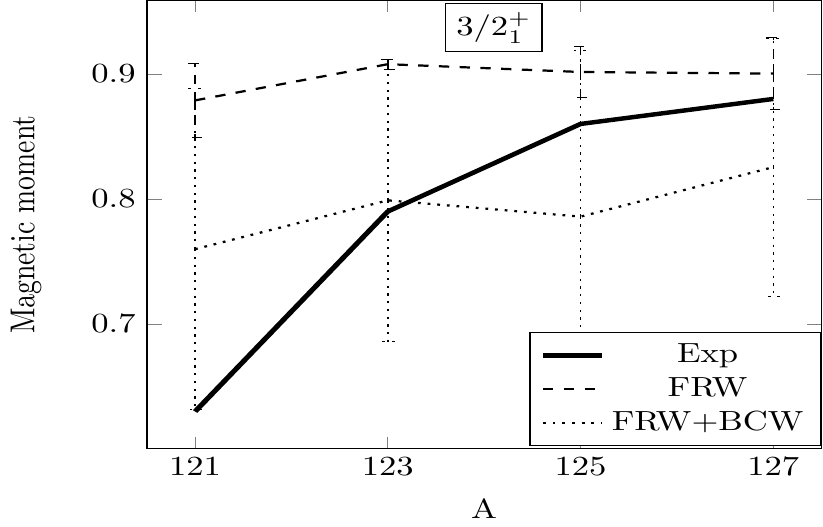}
  \caption{Magnetic moments in units $\mu_0$ of the first $11/2^-$,
    $3/2^+$ and $1/2^+$ states in the chain $^{111-127}$Cd for which
    experimental data is available \cite{2013_yordanov, stone}. The
    calculations are performed within the BCW+FRW (dotted line) and FRW
    (dashed line) versions of the model. The solid line represents the
    experimental values. The error bars give the uncertainty in
    evaluating the magnetic moments when varying the strength of the
    quadrupole-quadrupole interaction in a wide range of values. See
    text for details.}
  \label{fig:mu}
\end{figure}

For the evaluation of the electromagnetic moments it is of great
importance to know the values of the structural coefficients entering
into Eqs. \eqref{eq:a-qp-qp}, \eqref{eq:a-qp-ph} and
\eqref{eq:a-ph-ph}. In performing calculations for determining these
coefficients we retain only the quadrupole term from the multipole
expansion of the residual interaction between the nucleons' spatial
degrees of freedom and only the dipole term from the part describing
the residual interaction between the spacial
\eqref{eq:interaction_multipole-multipole} and spin
\eqref{eq:interaction_spin-multipole} degrees of freedom. In our
investigation the strengths of these interactions are free parameters
that are fixed by fitting on the experimental data. The mean field is
approximated, for simplicity, by the potential well of Woods-Saxon
form with the parameters determined by reproducing the nuclear binding
energies. The monopole pairing strengths $G_\tau$ are obtained to
match the odd-even mass differences in neighboring nuclei, as detailed
in Ref \cite{2008_mishev}. The strength of the isoscalar
quadrupole-quadrupole interaction $\kappa^{(2)}_{0}$ is adjusted so as
to reproduce the experimental spectrum of the low-lying states of each
individual odd-even nucleus. The dependence of the magnetic moment on
this parameter are shown in Fig. \ref{fig:mu}, varying it in a broad range
of values. The parameter $\kappa^{(2)}_{1}$ is calculated by using the
relation (conf. Ref. \cite{1983_vdovin}):

\begin{equation}
\kappa_{1}^{(\lambda)} = -0.2(2\lambda + 3) \kappa_{0}^{(\lambda)}.
\end{equation}

The isovector spin-multipole--spin-multipole 
interaction strength $\kappa^{(10)}_{1}$ is determined by the centroid of the
giant dipole magnetic resonance while the strength of the isoscalar
spin-multipole--spin-multipole interaction $\kappa^{(10)}_{0}$ plays
negligible role for the structure of the 1$^{+}$ giant resonance and
is set to 0.

\begin{table}[t]
  \centering
  \begin{tabular*}{\linewidth}{c | c c c |  c c| }
    & \multicolumn{3}{c|}{FRW+BCW}  & \multicolumn{2}{c|}{FRW} \\ \hline
    Isotope  & C(D) & E(F) & Component &    C(D) &  Component \\ \hline

    & 0.88&0.02& $\mathrm{\nu 1h_{11/2}}$                           & 0.98 & $\mathrm{\nu 1h_{11/2}}$    \\
    $^{121}$Cd  &-0.01&-0.42&$\mathrm{\nu 1h_{11/2}\otimes 2^+_1}$   &  0.08 & $\mathrm{\nu 2f_{7/2}\otimes2^+_1}$ \\
    &0.14&-0.11&$\mathrm{\nu 2f_{7/2}\otimes 2^+_1}$               & 0.08 & $\mathrm{\nu 1h_{11/2}\otimes 1^+_8}$ \\
    &0.08&0.00&$\mathrm{\nu 1h_{11/2}\otimes 1^+_8}$               & 0.07 & $\mathrm{\nu 1h_{11/2}\otimes 1^+_3}$  \\ \hline

    &0.87&-0.01&$\mathrm{\nu 1h_{11/2}}$                            &0.98 & $\mathrm{\nu 1h_{11/2}}$  \\
    $^{123}$Cd &-0.08&-0.41& $\mathrm{\nu 1h_{11/2}\otimes 2^+_1}$   &0.10 & $\mathrm{\nu 1h_{11/2}\otimes 1^+_8}$ \\
      &0.14&-0.11&$\mathrm{\nu 2f_{7/2}\otimes 2^+_1}$               &0.07&$\mathrm{\nu 2f_{7/2}\otimes 2^+_1}$ \\
      &0.1&0.00&$\mathrm{\nu 1h_{11/2}\otimes 1^+_8}$               &-0.06 & $\mathrm{\nu 1h_{9/2}\otimes 1^+_8}$  \\ \hline

     &0.91&-0.03&$\mathrm{\nu 1h_{11/2}}$                           &0.98 & $\mathrm{\nu 1h_{11/2}}$  \\
    $^{125}$Cd &-0.15&-0.30&$\mathrm{\nu 1h_{11/2}\otimes 2^+_1}$      & -0.13 & $\mathrm{\nu 1h_{11/2}\otimes 2^+_1}$ \\
     & 0.09&-0.10&$\mathrm{\nu 2f_{7/2}\otimes 2^+_1}$               & 0.10 & $\mathrm{\nu 1h_{11/2}\otimes 1^+_8}$ \\
     &0.09&0.00&$\mathrm{\nu 1h_{11/2}\otimes 1^+_8}$               & 0.07 & $\mathrm{\nu 2f_{7/2} \otimes 2^+_1}$  \\ \hline

     &0.92&-0.05&$\mathrm{\nu 1h_{11/2}}$                           & 0.97 & $\mathrm{\nu 1h_{11/2}}$   \\
    $^{127}$Cd&-0.22&-0.23&$\mathrm{\nu 1h_{11/2}\otimes 2^+_1}$       &-0.20 & $\mathrm{\nu 1h_{11/2}\otimes 2^+_1}$ \\
    &0.10&0.00&$\mathrm{\nu 1h_{11/2}\otimes 1^+_8}$                &0.10  &$\mathrm{\nu 1h_{11/2}\otimes 1^+_8}$  \\
     &0.06&-0.08&$\mathrm{\nu 2f_{7/2}\otimes 2^+_1}$                &0.05 &$\mathrm{\nu 2f_{7/2}\otimes 2^+_1}$  \\
  \end{tabular*}
  \caption{Major components of the 11/2$^{-}_{1}$ state in $^{121}$Cd, $^{123}$Cd, $^{125}$Cd and $^{127}$Cd calculated using both forward (third column) and forward+backward (second column) amplitudes in the wave functions }
  \label{tbl:structure}
\end{table}

The structural compositions of the wave functions of the $11/2^-_1$
states in the studied isotopes is a result of the interplay
between the neutron subshell $1h_{11/2}$ and the
quasiparticle$\times$phonon configurations in the ground and in the
excited states. The calculated components of these wave functions in
versions of the model including backward amplitudes (FRW+BCW for
short) and disregarding them (abbreviated by FRW) are listed in Table
\ref{tbl:structure}. As seen in this table, by allowing
quasiparticle$\otimes$phonon configurations in the ground state, one
obtains an increased fragmentation of the $1h_{11/2}$ quasiparticle
strength (conf. Refs. \cite{1993_Sluys,2008_mishev}) which causes a
reduction in the contribution $x_{qp-qp}$ (see Eq.\eqref{eq:x-qp-qp})
to the quantities of interest. In the FRW+BCW model version the
largest part of this strength is transferred to the $\nu
1h_{11/2}\otimes 2^+_1$ admixture in the ground state of the even-even
core which interacts most intensely with the last quasiparticle. The
main part of the strength of this interaction is given by

 \begin{align}
   \label{eq:W}
   W \left( {Jj\lambda 1} \right) & = \left\langle {\left| \alpha^{\dagger}_{JM} H P_{j\lambda i}^{\dagger}(JM)\right|} \right\rangle = \nonumber 
\\
   & = - \frac{1}
   {4}\frac{{\pi _\lambda }} {{\pi _{J }
     }}\sum\limits_{\tau _0 } {{\CA}_{\tau _0 } \left( {\lambda i 1}
     \right)\varphi _{J j }^{\lambda 1} } \hfill
 \end{align}

 , where $\varphi _{J j }^{\lambda i}$ are the phonons' backward
 amplitudes and ${\CA}_{\tau _0 } \left( {\lambda i i'} \right)$ is
 defined in \cite{2008_mishev}.

 \begin{figure}[t]
   \includegraphics{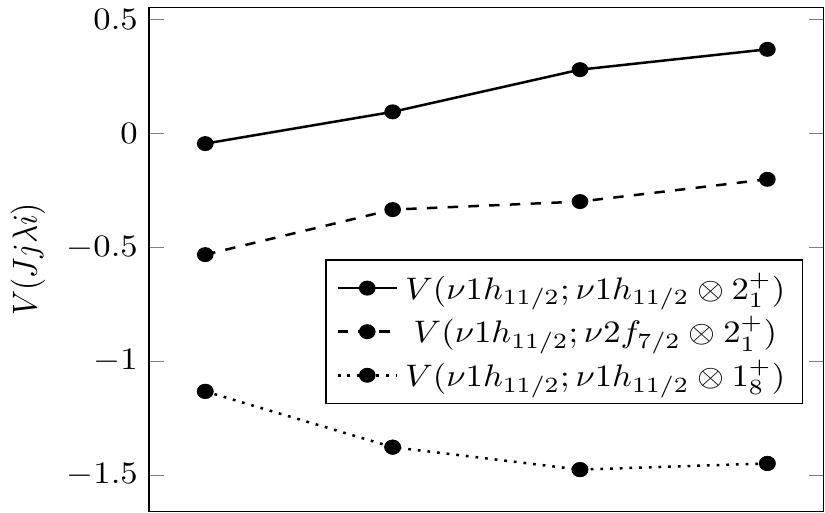}
   \includegraphics{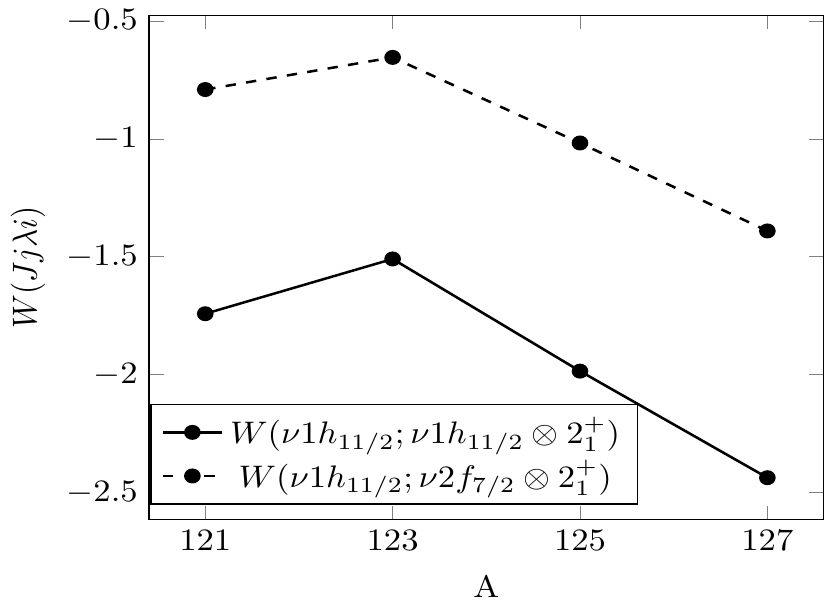}
   \caption{Evolution of select interaction vertices in the forward $V(Jj\lambda i)$ and backward $W(Jj\lambda i)$ directions with the mass number $A$.}
   \label{fig:V}
 \end{figure}

 As seen from Eq.\eqref{eq:W}, the only two-quasiparticle state which
 influences this interaction strength is the one which bears nucleons
 from shells designated by the quantum numbers of the participating
 non-collectivized quasiparticles, namely $J$ and $j$. The high
 amplitude $\varphi_{Jj}^{\lambda i}$ (between 0.3 and 0.5 in
 different isotopes) of annihilating the two-quasineutron state $[\nu
 1h_{11/2} \otimes \nu 1h_{11/2}]_2$ in the ground state for the
 formation of the $2_{1}^{+}$ phonon, explains the enhanced magnitude
 of the interaction between the $\nu 1h_{11/2}$ and $\nu
 1h_{11/2}\otimes 2_{1}^{+}$ states, which varies between -1.5 MeV to
 -2.5 MeV along the isotope chain as in Fig. \ref{fig:V}. The
 contribution from the configuration $[\nu 2f_{7/2}\otimes
 2_{1}^{+}]_{11/2}$ to the structure of the $11/2^{-}_{1}$ state is
 the second largest. The reason for such a high rank of this component
 is the considerable (of the order of 0.13) $2_1^+$ phonon amplitude
 for annihilating $[\nu 1h_{11/2}\otimes \nu 2f_{7/2}]_{2}$ state
 residing in the ground state of the neighboring even-even nucleus.

On the other hand, if the particle-vibration interaction in the ground
state is not taken into account, then the structure of the
$11/2^{-}_{1}$ states change significantly from one nucleus to
another, as seen from Table \ref{tbl:structure}. The reason for the
reordering of the components is the interaction between the last
particle and different excited quasiparticles$\otimes$phonon
configurations, which is quantified by the vertex:

 \begin{align}
   \label{eq:V}
   & V \left( {Jj\lambda i} \right) = - \left\langle {\left| { { \alpha _{J M } H P_{j\lambda i }^{ + }
         \left( {J M } \right)} } \right|} \right\rangle = \nonumber \\
   & = \begin{cases}
     \frac{1}
          {{\sqrt 2 }}\frac{\pi_{\lambda}}{\pi_{J}} \frac{
            f_{Jj}^{(\lambda )} v_{Jj}^{(-)}}{\sqrt{Y^{\lambda i}}} , \mbox{for } \lambda^\pi = 2^+  \\
     \frac{1}
          {{\sqrt 2 }}\frac{\pi_{\lambda}}{\pi_{J}} \frac{
            f_{Jj}^{(\lambda L)} v_{Jj}^{(+)}}{\sqrt{Y^{\lambda i}}} , \mbox{for } \lambda^\pi = 1^+,
   \end{cases}
 \end{align}

 where $v_{Jj}^{(\pm)} = u_{J}u_{j} \pm v_{J}v_{j}$ with $u_j$ and
 $v_j$ being the pairing occupation numbers for the level $j$.
 
 The trends for these interaction vertices, imposed by the
 relationship in Eq.\eqref{eq:V}, when changing the neutron number are
 complex because of their dependence on both the pairing properties of
 the non-collectivized particles and the degree of collectivity of
 the participating phonon. For instance, the key to understanding the
 trends in $V \left( {\nu 1h_{11/2};\nu 1h_{11/2}\otimes 2_{1}^{+}} \right)$,
 plotted in Fig. \ref{fig:V}, are the changes in the values of $v_{\nu
   1h_{11/2}, \nu 1h_{11/2} }^{(-)}$ which drop down from 0.4619 in
 $^{117}$Cd passing through 0.05 in $^{121}$Cd and reaching -0.530 in
 $^{127}$Cd. Analogously, the line for the vertex $V \left( {\nu
     1h_{11/2};\nu 2f_{7/2}\otimes 2_{1}^{+}} \right)$ is explained by
 the attenuation of the quantity $v^{-}_{\nu 1h_{11/2},\nu 2f_{7/2}}$
 from 0.61 in $^{121}$Cd to 0.36 in $^{127}$Cd. In contrast, the
 vertex $V \left( {\nu 1h_{11/2};\nu 1h_{11/2} \otimes 1^{+}_{8}}
 \right)$ does not depend on the pairing directly since $v_{\nu
   1h_{11/2}, \nu 1h_{11/2}}^{(+)}=1$ and its evolution over the
 isotope chain is driven by the collective properties of the $1^+_8$
 state in the respective even-even core.

 The first order core polarization is manifested by the significant
 contribution to the magnetic moment of the
 quasiparticle$\otimes$1$^+$-phonon states. The structure of the
 $1^{+}$ states has a simple form and is represented by a mixture of
 coupled quasiparticle states belonging to spin-orbit doublets.
 Although the contribution to the odd-even nucleus wave function
 coming from the configurations including $1^{+}$ states is small,
 its importance to describing the magnetic moments is vital, as it
 will be discussed in the following section. The 1$^+$ state which
 mostly contributes  to the structure of the first 11/2- states in
 the odd-even cadmium isotopes is the one acquiring the highest degree
 of collectivity and having an energy in the region of 13 MeV. Its
 structure is dominated by the configuration $[\nu 1h_{11/2}\otimes
 \nu 1h_{9/2}]_{1}$. If the higher order correction, given by
 Eq.\eqref{eq:a-ph-ph}, could be neglected, then the coupling of the
 last nucleon with the 1$^+$ states might be treated by following the
 classical approaches by Blin-Stoyle \cite{1954_blin-stoyle} and Arima
 and Horie \cite{1954_arima}, which are based on the smallness of the
 mixing coefficients. In our calculations we confirm that the
 perturbing configuration from Ref.\cite{1954_blin-stoyle}, which
 applied to the 11/2- state in the cadmium isotopes is
 $\left[[\nu 1h_{11/2}\otimes \nu 1h_{9/2}]_{1}\otimes
   \nu 1h_{11/2}\right]_{11/2}$, indeed contributes the most to the
 magnetic moment if the backward amplitudes are not taken into
 account. However, the inclusion of a single complex configuration to
 the wave function is not adequate for describing the magnetic moments
 because the coupling with excitation modes of the even-even core
 having higher angular momenta is crucial to gain a more realistic
 picture of the wave function.

\subsection{Magnetic moment}
\label{sec:magnetic_moments}

\begin{figure*}
  \includegraphics{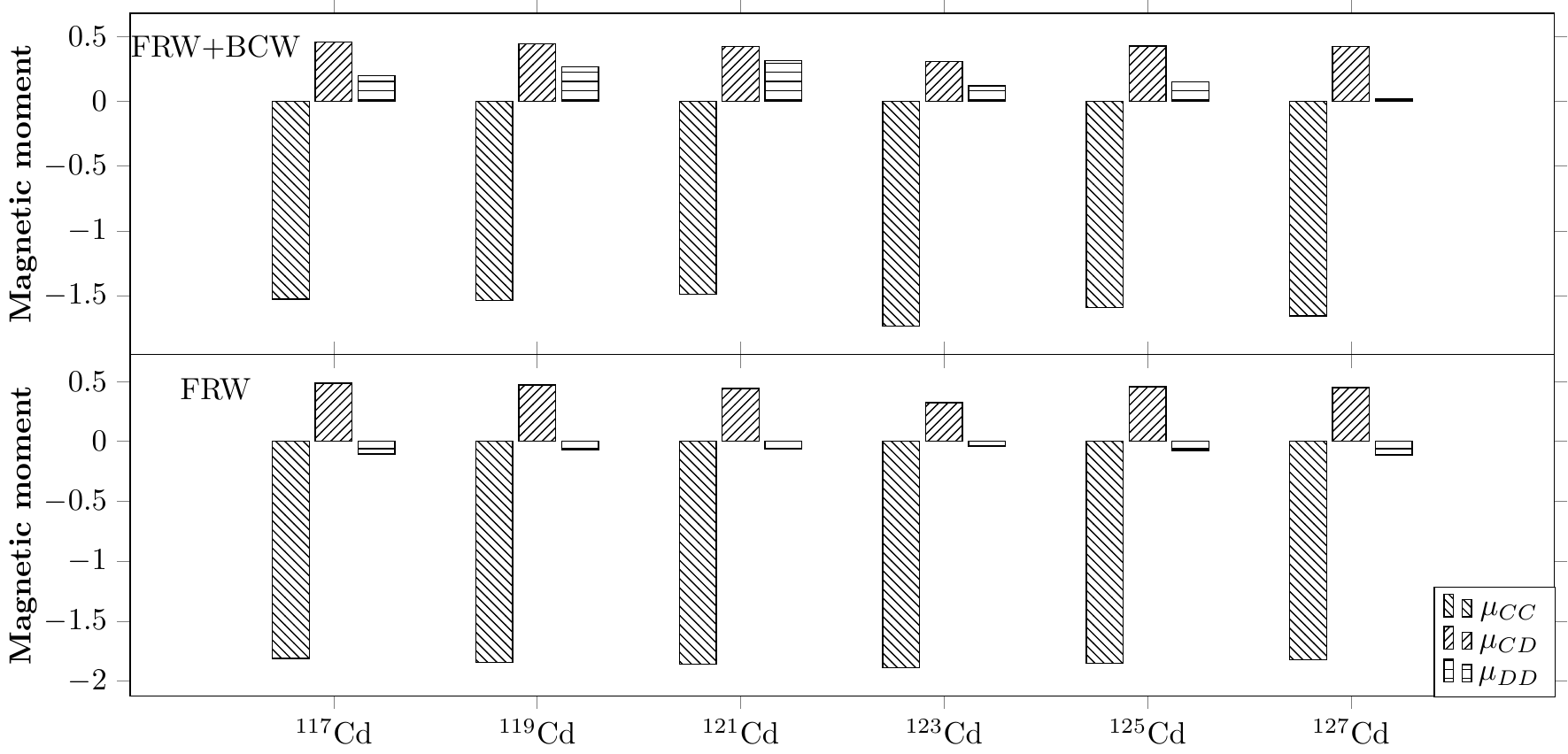}
  \caption{Contribution to the magnetic moment of the 11/2- state in the series of isotopes $^{117-127}$Cd}
  \label{fig:mu_contrib}
\end{figure*}

The particularities in the structure of the odd-A isotopes discussed
in the previous section determine the deviations of the magnetic
moment of the $11/2_{1}^{+}, 3/2^{+}_{1}$ and $1/2^{+}_{1}$ states from
their single-particle estimates. The results from the calculations
performed by using the featured QPM versions are plotted in
Fig. \ref{fig:mu} and are compared to the experimental values. The
dotted and dashed lines in this figure depict the most important
achievement of this work - the finding that the interaction between
the last quasiparticle and the ground-state phonon admixtures produces
configurations which contribute significantly to the magnetic moment
of odd-A nuclei and reveal a potential for reproducing their
experimental values, which proves impossible if this interaction is
neglected. The importance of each contribution from
Eqs. \eqref{eq:a-qp-qp}-\eqref{eq:a-ph-ph} to the magnetic dipole
moment is visualized in Fig.\ref{fig:mu_contrib}. The enhanced
fragmentation due to the quasiparticle-phonon interaction in the
ground state leads to systematically shrinked values of the single
quasiparticle contribution $\mu_{qp-qp}$ and to an increase in the
quasiparticle-phonon contribution $\mu_{qp-ph}$ leading to an overall
decrease in the magnitude of the magnetic moment. The escalation of
the magnetic transitions between different quasiparticle$\otimes$phonon
configurations, given by $\mu_{ph-ph}$, is due to configurations
involving a quadrupole phonon, of which $\nu 1h_{11/2}\otimes 2_{1}^{+}$
plays the most important role.  It is worth noting that because of the
weakened coupling between the quasiparticles and the quadrupole
phonons in the core's ground state near the neutron shell closure, the
quantity $\mu_{ph-ph}$ tends to diminish while $\mu_{qp-qp}$ remain
almost unchanged along the isotope chain. This interaction, however,
leaves the first order core polarization term $\mu_{qp-ph}$,
describing the magnetic transitions between quasiparticle and
quasiparticle$\otimes 1^+$-phonon states, virtually unaffected because
the latter configurations represent a negligible part in the
$11/2^-_1$ state mixture.

However, despite its capacity of reaching the experimental values,
this theoretical development suffers from the shortcoming
(conf. \cite{1993_Sluys, 2008_mishev}) that the residual interaction
strength, which yields results that are of sound agreement with the
odd-A experimental data, generates substantially less collective
$2_1^+$ states in the even-even cores than the ones implied from the
data for the neighboring even-even nuclei. The origin of this
inconsistency is the set of approximation techniques embedded in the
considered QPM versions, namely, the BCS and RPA, which tend to
overestimate the degree of correlations in the ground state for
open-shell nuclei. One path to healing this problem is to apply the
more consistent Extended RPA \cite{erpa}, or use a tractable
method based on the variational principle for the ground state, as in
\cite{2013_mishev}.

\section{Conclusions}
\label{sec:outlook}

The magnetic dipole moments of the low-lying states in odd-A Cd nuclei
are found to be significantly affected by the correlations in the
ground state. The obtained corrections allow one to reproduce the
experimental values in open-shell nuclei, which proves impossible if
the existence of the quasiparticle$\otimes$phonon configurations in
the ground states of even-even is ignored. Despite the reported
improvements, the calculations based on this version of the model
exhibit a very high sensitivity on interaction parameters which
limit its predictive power, and pertinent work in this direction is
ongoing.

\begin{acknowledgments}
  S. Mishev would like to thank Dr. D. Yordanov for useful remarks and
  discussion. S. Mishev acknowledges the financial support from the
  Bulgarian National Science Fund under contract No. DFNI-E02/6.
\end{acknowledgments}

\end{document}